\documentclass[pre,aps,twocolumn,showpacs]{revtex4}
\usepackage{dcolumn}
\usepackage{graphicx}
\usepackage{graphicx,amssymb,amsmath}

\usepackage{natbib}
\usepackage{bm}

\begin{document}
\title{The nonlinear elasticity of an $\alpha$-helical polypeptide}

\author{Buddhapriya Chakrabarti and Alex J. Levine}

\affiliation{Department of Physics, University of Massachusetts.\\ Amherst, MA 01003}

\date{\today}

\begin{abstract}
We study a minimal extension of the worm-like chain to describe polypeptides having alpha-helical secondary structure. In this model presence/absence of secondary structure enters as a scalar variable that controls the local chain bending modulus. Using this model we compute the extensional compliance of an alpha-helix under tensile stress, the bending compliance of the molecule under externally imposed torques, and the nonlinear interaction of such torques and forces on the molecule. We find that, due to coupling of the ``internal'' secondary structure variables to the conformational degrees of freedom of the polymer, the molecule has a highly nonlinear response to applied stress and force couples. In particular we demonstrate a sharp lengthening transition under applied force and a buckling transition under applied torque. Finally, we speculate that the inherent bistability of the molecule may underlie protein conformational change \emph{in vivo}.
\end{abstract}

\pacs{87.10.+e,87.14.Ee,82.35.Lr}

\maketitle

\section{Introduction}

The study of the mechanical properties of individual biopolymers
serves as an important laboratory to probe polymer physics at the
length scale of a single chain and further elucidates the
biological processes in which these molecules take
part\cite{Austin:97,Strick:03}. For example the biologically
fundamental processes of DNA replication, transcription, the
regulation of transcription rely on the DNA protein interactions
involving the mechanical deformation and microstructural
modification of DNA both at long length scales as well as at the
scale of individual base pairs. Recent advances in the
experimental manipulation of individual biological macromolecules
has opened a new window on these processes and allows for the
direct quantitative test our understanding of the mechanical
properties of these macromolecules in thermal equilibrium. These
single-molecule manipulation experiments have probed the
mechanical properties of not only
DNA\cite{Smith:92,Perkins:94,Perkins:95,Cluzel:96,Perkins:97,Bustamante:03,Strick:03}
but also a variety of biologically important macromolecules
including polysaccharrides\cite{Rief:97,Li:98}, and giant proteins such as titin \cite{Tskhovrebova:97,Kellermayer:97,Rief:97a} and 
tenascin \cite{Oberhauser:98}. The better theoretical understanding of protein mechanics will enhance the interpretation of protein force spectroscopy, which may even shed light on protein folding pathways \cite{Carrion-Vazquez:99} although this latter point appears to be somewhat controversial\cite{Best:01,Fowler:02}.  Regardless, understanding the mechanical properties of proteins is fundamental to elucidating the allosteric or conformation changes that many proteins undergo as part of their biological activity\cite{Frauenfelder:91,Eaton:99,James:03,Goh:04}. 

These last examples demonstrate the feasibility of the direct
mechanical manipulation of single proteins. However, modelling the mechanical properties of these atypically large proteins or, for that matter,  any entire protein is a daunting task since such molecules have complex structures that result from a number of local and nonlocal interactions along the polymer chain. In order to make quantitative progress in the interpretation of these protein manipulation experiments it appears to be useful to first understand in more detail the mechanical properties of simpler polypeptide-based structures. A natural candidate for such a simpler structure is a protein subdomain of one secondary
structure. Here too there is experimental input: single molecule force spectroscopy via AFM has been used to directly probe alpha-helical polypeptides\cite{Lantz:99} as well as synthetic polymer chains with a local helical structure such as PEG \cite{Oesterhelt:99}.

In order to understand the mechanical properties of proteins in general and alpha-helical polypeptides in
particular it is necessary to develop a new minimalistic model
that incorporates both the conformational fluctuations of the
polymer backbone and localized structural transitions of the
constituent monomers. In other words it is necessary to augment simple models of the statistical mechanics of the peptide backbone with new terms to account for the presence of secondary structure along the chain. Furthermore, we must allow for interaction between the degree of local secondary structure and the conformational degrees of freedom of the polypetide backbone.

In this paper we examine the predicted mechanical properties of
such a minimalistic model of an alpha-helical polypeptide in which we allow the interaction of the local secondary structure of the chain with its conformational degrees of freedom. We treat the local presence or absence of secondary structure as a two-state (Ising-like) variable along the chain backbone, which is itself described by a set of local tangent vectors to the chain. In order to make this simplification, we coarse-grain the polymer so that each independent monomer can be unambiguously assigned a state of secondary structure. This requires us to consider a model comprised of coarse-grained chain segments (\emph{i.e.} monomers) each consisting of about three amino acids.  The interaction between the internal, secondary structure variables and the conformation of the polymer chain as described by the set of backbone tangent vectors is effected by the presence of a bending modulus of the backbone whose value depends on the local state of secondary structure.  When these three amino acids making up one model monomer adopt a local configuration consistent with alpha-helical secondary structure, the hydrogen bonding between these amino acids\cite{Brandon:99} renders that segment of the chain significantly stiffer than the same polymer without the locally ordered secondary structure. Thus the bending energy associated with the local change in the backbone tangent vectors is higher in regions having alpha-helical structure than in regions locally adopting a random coil configuration. Similar models can and have been applied to study the mechanical properties of DNA and have been discussed in the current context as well\cite{Buhot:00,Buhot:02,Storm:03}.

To qualitatively characterize our results presented below, we note that, due to the presence of the internal state variables representing secondary structure along the chain and their control over the local chain bending modulus, the alpha-helix is predicted to have a highly nonlinear response to both bending torques and to extensional forces.
Under small externally applied torques, the molecule will deform so that its thermally averaged chain contour takes the form of the arc of a circle and the torque necessary for bending the molecule through a given angle grows linearly with that angle. The molecule deforms roughly as a flexible rod. At a critical torque, however, the secondary structure of the molecule is locally disrupted producing a small length of the backbone with a much softer bending modulus. The total curvature that had been uniformly distributed along the backbone becomes localized in the anomalously soft region produced by the disruption of the secondary structure and the torque required to enforce the curvature of the molecule drops precipitously.

The long range goals of this
sort of modelling go beyond the interpretation of the
emerging experiments on the mechanical properties of
polypeptides having alpha-helical secondary structure.
By understanding the mechanical properties of the constituent
elements of a protein it should be possible to develop a
lower dimensional representation of protein mechanics. In place of the atomic coordinates of the backbone carbons and the positions of the various amino acid residues, one may describe protein domains (having definite secondary structure) as a space curve having some nonlinear extensional and bending compliances that may be computed in terms of a few
energy scales determined either from experiment or
simulation. Using three-dimensional protein structural data
and such a nonlinear elastic model for each structural element
of definite secondary structure, one can attempt to build
 mechanical models of entire proteins that, due to their highly reduced number of degrees of freedom are more tractable for numerical investigation than those based on all atom simulations. 

From the study of those models one may be able to extract low energy conformational pathways and thus make predictions regarding protein allostery.  For example, from a combination of native-state protein structural data and the calculated nonlinear elastic properties of alpha-helical protein domains it may thus be possible to predict the mechanical properties of the alpha-helical coiled coil region in myosin II \cite{Li:03} or alpha-helix rich proteins such as spectrin  as probed by mechanical unfolding experiments \cite{Law:03,Law:03a}. Further
data on protein conformational change is  available from numerical simulation\cite{Choi:03,Schulten:00}.

The remainder of the paper is organized as follows. In section
\ref{model} we introduce the alpha-helix Hamiltonian based on a
combination of the worm-like chain and the helix/coil model. Using this model we calculate the response of the chain to bending torques in section \ref{bending}. We then take up the problem of the extensional compliance of the alpha-helix in section \ref{stretching} before summarizing the results and discussing possible experimental tests of the theory in section
\ref{conclusions}.

\section{The Helix-coil worm-like chain}
\label{model}

The worm-like chain (WLC)\cite{Kratky:49,Fisher:63} is the
fundamental coarse-grained model for a polymer at length scales
shorter than its thermal persistence length. This model describes
the single-chain polymer statistics in terms of a quadratic
Hamiltonian that associates an energy cost with chain curvature by introducing a bending modulus $\kappa$. In terms of a discretized chain model described by the set of monomeric tangent vectors
$\hat{t}_i$, $i = 0, \ldots, N-1$ with $N$ the degree of
polymerization, the WLC Hamiltonian may be written as
\begin{equation}
\label{WLC-Hamiltonian} H_{\rm WLC} = \kappa \sum_{i=0}^{N-1}
\left[ 1 - \left(\hat{t}_i \cdot \hat{t}_{i+1} \right) \right].
\end{equation}
The effect of this bending energy is to enhance the statistical
weight of straight chain configurations on a length scale of
$\kappa$ monomers equal to the thermal persistence length. Here and throughout this paper we take $k_{\rm B} T = 1$.  It may be easily checked that this length is equal to the arc length (measured in  monomer lengths $\gamma$) of the polymer chain over which the chain tangent vectors thermally decorrelate. At length scales much longer than $\kappa \gamma$ the effect of this bending energy is minimal and the equilibrium statistics of the polymer become controlled by a combination of intrachain collisions and chain configurational entropy \cite{deGennes:79} with a renormalized Kuhn length.  In theta solvent one finds in the limit of very long worm-like chains  the radius of gyration to be given by $\langle R_g^2
\rangle = 2 \kappa \gamma^2 N$ where $\gamma$ is the monomeric length.

The single-chain response to externally applied tension $F$ has
also been exhaustively researched within the WLC description of
the polymer \cite{Marko:95, MacKintosh:95,Lamura:01}. The fundamental result of this work is that one may determine the extensional compliance of the molecule as a function of applied force. This compliance is defined as $\partial \langle L \rangle /\partial F$, the derivative of the equilibrium chain length $\langle L \rangle$
with respect to the applied force $F$. At low forces it is
essentially constant reflecting the standard, quadratic reduction
in chain configurational entropy associated with long, flexible
polymers. At high forces, however, the compliance goes to zero as
$F^{-3/2}$ due to the fact that the chain has a finite length at even arbitrarily high forces. The characteristic form of the approach of the compliance to zero in the high force limit is controlled by the pulling out of small, transverse thermal fluctuations of the chain and thereby recovering the arc length stored in them to increase $\langle L \rangle$.

While the WLC is a highly successful model to describe the force
extension properties of a number of biopolymers such as DNA, it is clear that it is not sufficient to properly describe these
molecules under large enough tensions. At larger tensile stresses, details of the internal structure of the molecule become important for the understanding of conformational properties of the molecule. For example, under large enough stresses the double-helix structure of DNA (B-DNA) can be unwound allowing each monomer to lengthen by a factor of about 1.85 \cite{Smith:95}.  To account for such (two-state) internal degrees of freedom along the chain, workers have employed the helix/coil (HC) model\cite{Poland:70}. This model has been used to study a class of protein conformational
transitions\cite{Birshtein:66,Bloomfield:99} in solution and under tension\cite{Tamashiro:01}.

\begin{figure}
\includegraphics[width=7cm]{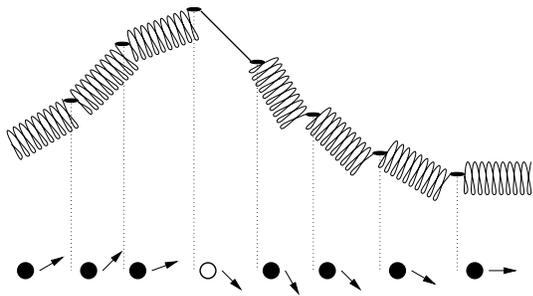}
\caption{\label{finalschem} Schematic figure of an alpha-helical polypeptide and its schematic representation in terms of the Ising-like secondary structure variables (open circles for random coil segments and filled ones for alpha-helical ones) and the tangent vectors to the segments of the chain (denoted by arrows).}
\end{figure}

The HC model Hamiltonian, which is used to model these structural
transitions can be reduced to it simplest form by assuming that
the local structure of the chain can be described by a set of
two-state variables $s_i = \pm 1$, $ i = 0, \ldots, N$. For the
alpha-helical chains of current interest we regard these two
states as the local conformation of the monomer in its native,
alpha-helical state ($s = +1$) and in a disordered, random coil
state ($s = -1$).  The statistics of this set of two-state
variable is controlled by the Hamiltonian:
\begin{equation}
\label{hc-hamiltonian}
 H_{HC} = \epsilon_{\rm w}/2
\sum^{N-1}_{i=0} (1 - s_i s_{i+1}) - h/2 \sum^{N}_{i=0} (s_i - 1).
\end{equation}
It is immediately clear that above Hamiltonian also describes a
one dimensional, ferromagnetic Ising chain. The interpretation
here is somewhat different. The energy $h$ playing the role of an
external magnetic field in the Ising system now represents the
free energy cost per monomer to be in the non-native (\emph{i.e.}
random-coil) state. This term is thus controlled by a combination
of the chemistry of the monomeric residues and solvent quality;
its calculation from fundamental solution chemistry is beyond the
scope of the current work, however we will attempt to estimate its magnitude based on experiment. Clearly this constant is at least of order unity since the protein domain under investigation is assumed to have an alpha-helical secondary structure in thermal equilibrium.

The first term in Eq.~\ref{hc-hamiltonian} plays the role of the
nearest neighbor ferromagnetic coupling in the Ising
interpretation of the Hamiltonian. In its current interpretation,
$\epsilon_{\rm w}$ is free energy cost of a domain wall in the
sequence of helix ($s = +1$) and random coil ($s = -1$) sites.  In the helix/coil literature it is also referred to as the natural logarithm of the ``chain-cooperativity'' parameter. By adopting a native state configuration, a monomer presents hydrogen bonding sites to its neighbors. If those neighbors are also in their native state, these hydrogen bonds further lower the free energy of the system via this nearest neighbor cooperative effect. If, however, one of the neighboring monomers of an alpha helical monomer is in its random-coil state, such hydrogen bonding is not possible and the total free energy of this domain wall configuration is larger than simply the free energy cost for one monomer being in the non-native configuration, $h$. 

Finally, we note that since there are
multiple hydrogen bonds per turn of the alpha helix, it might be
reasonable to describe the local secondary structure by a
$q$-state discrete variable where $q > 2$. Such an analysis
changes Eq.~\ref{hc-hamiltonian} into an equally tractable
one-dimensional $q$-state potts model, but introduces additional
unknown parameters. It is thus inconsistent with our goal of exploring a minimal model that incorporates secondary structure.

The coupling of the secondary structure variables to the WLC tangent vectors is affected by introducing a bending stiffness in the WLC Hamiltonian that depends on the local degree of secondary structure. We choose
\begin{equation}
\label{kappa-def} \kappa(s) = \left\{ \begin{array}{ll}
                        \kappa_> & \mbox{if $s = + 1$} \\
                        \kappa_< & \mbox{if $s=-1$}
                   \end{array}
            \right.  .
\end{equation}
Due to the hydrogen bonding between turns of the alpha-helix, it
is reasonable to expect that $\kappa_>$ the bending modulus in the native state is significantly larger than $\kappa_<$ the bending modulus of the chain in the non-native, disordered state. We return to the question of determining physically reasonable estimates of these quantities in the conclusions. By introducing Eq.~\ref{kappa-def} and combining the Hamiltonians in Eqs.~\ref{WLC-Hamiltonian},\ref{hc-hamiltonian} we write the full Hamiltonian for the coupled system of secondary structure
variables and chain tangent vectors as the helix/coil worm-like
chain Hamiltonian (HCWLC): 
\begin{eqnarray}
 H = \epsilon_{\rm w}/2 \sum^{N-1}_{i=0} (1 -
s_i s_{i+1}) - h/2 \sum^{N}_{i=0} (s_i - 1)+ \nonumber \\
+ \sum^{N-1}_{i=0} \kappa(s_{i}) \left[ 1 - \left(\hat{t}_i \cdot
\hat{t}_{i+1} \right) \right]. \label{HCWLC-hamiltonian}
\end{eqnarray}
We note that the system described by Eq.~\ref{HCWLC-hamiltonian}
may be looked at as two intercalated Heisenberg ($\hat{t}_i$) and
Ising ($s_i$) magnetic systems. The nearest neighbor coupling of
the Heisenberg (chain tangent) vector , however, depends on the
value of the Ising (secondary structure) variable between them. A
pictorial representation of the system along these lines is shown
in figure \ref{finalschem}. The full Hamiltonian given by
Eq.~\ref{HCWLC-hamiltonian} has four constants with dimensions of
energy: $\kappa_>, \kappa_<, h, \epsilon_{\rm w}$ that can be fit
from experiment. We return to this point in our conclusions.
Finally, we note that we have disregarded the twist degree of
freedom of the molecule. Such twist degrees of freedom and the
coupling of twisting and stretching modes of these chiral
molecules have been explored particularly with regard to the
mechanical properties of DNA \cite{Kamien:97,Ohern:98}. This
extension of the basic model will be explored in future work.

\section{Mechanical Properties}

We now turn to the exploration of the mechanical properties of the polymer described by the HCWLC Hamiltonian. We study this problem in different ways. First, we consider the response of the
molecule to externally applied torques by examining the torque
required in thermal equilibrium to enforce a given angular
deviation between the first and last chain tangent vectors.
Second, we study the force--extension relations for this molecule
by calculating the projection of the chains's mean end-to-end distance along the direction of an applied force as a function of the magnitude of that force. We study this extensional compliance for two different cases. In one case we assume that the end tangents of the chain remain unconstrained. In the second,  we explore the effect of applied torque on the extensional compliance of the molecule by first constraining the end tangents vectors and then applying an extensional force. All these calculations are conceivable as individual, single molecule experiments through the use of \emph{e.g.} optical and magnetic traps. By understanding the dependence of the extensional compliance on the curvature of the molecule, one may gain insight into the mechanical properties of alpha-helical domains of proteins that are similarly constrained in the protein's native state.

\subsection{Bending}
\label{bending}

To consider the bending response of the chain to applied torques
in thermal equilibrium we first express the restricted partition
function of the system subject to the constraint that the chain
tangent deflects by a fixed angle $\psi$ over its total arc
length. It is reasonable to suppose that the chain bends in the
plane defined by first and last chain tangent that are being
constrained; to simplify the calculation we assume that we may
examine the problem in this two-dimensional subspace. In that case
the chain tangents are each equivalent to single angle:
$\hat{t}_{i+1} \cdot \hat{t}_i  \longrightarrow \cos(\theta_{i+1}-
\theta_i)$ and we write the restricted partition function as
\begin{equation}
\label{bending-partition}
 Z(\psi) = \prod^{N}_{j=0} \sum_{s_{j} =
\pm 1} \int \prod^{N}_{i=0} d \theta_{i} e^{-H} \delta(\theta_{0})
\delta(\theta_{N}-\psi),
\end{equation}
where the delta functions enforce the constraints on the initial
and final chain tangents.  The Hamiltonian appearing above is
given by Eq.~\ref{HCWLC-hamiltonian} and is a function of all the
secondary structure variables and backbone tangents.

To evaluate the partition function it is useful to observe the
formal equivalence of Eq.~\ref{bending-partition} to the imaginary
time propagator of a quantum particle on the unit circle
\cite{Kleinert:90}.  The partition function above is an
imaginary--time sliced  path integral
representation of the transition amplitude for the particle to
start at angle $\theta_0 = 0$ and end at angle $\theta_N = \psi$
in $N$ time slices.  The imaginary time evolution operator is
simply the exponentiated Hamiltonian appearing in
Eq.~\ref{bending-partition}. The quantum analogy is somewhat
complicated by the presence of the secondary structure variables;
for the current problem the fictitious quantum particle has a
two--level ``internal'' variable similar to the spin states of a
spin $1/2$ particle. The state of this particle in the angle representation take the form: $| s, \theta \rangle$ and
Eq.~\ref{bending-partition} can be recast in the form
\begin{equation}
\label{Z-quantum}
 Z(\psi) = \sum_{s_0, s_N=\pm 1} \eta_{s_N}
\langle s_0, \theta_0=0 | T^{N} |s_N, \theta_N=\psi \rangle,
\end{equation}
where $T$ is the single--step imaginary time evolution operator
and $\eta_s = e^{-h} \delta_{s,-1} $ is a factor needed to
correct the statistical weight of finding the last chain monomer
in the disordered, non-native state.  The remaining sum is
over the starting and ending secondary structure of the chain.

To make progress it is useful to work in terms of the integral
angular momentum variables $m_i$ conjugate to the angles
$\theta_i$. In this angular momentum representation of the problem
we may expand the angle eigenstates as
\begin{equation}
\label{m-expansion}
| s, \theta \rangle = \vert s \rangle \otimes \frac{1}{\sqrt{2 \pi}}
\sum_{m = - \infty}^\infty  e^{- i m \theta}.
\end{equation}
In this momentum representation the time evolution operator is diagonal, \emph{i.e.} it connects states of the same $m$ only:
\begin{eqnarray}
\langle m s | T | m^\prime s^\prime \rangle = \nonumber \\
\delta_{m, m^\prime}
    \begin{pmatrix}
     e^{-\kappa_{>}} I_{m}[\kappa_{>}]      &  e^{-\epsilon_{\rm w}-\kappa_{>}} I_{m}[\kappa_{>}] \\
     e^{-\epsilon_{\rm w}-h-\kappa_{<}} I_{m}[\kappa_{<}]        & e^{-h-\kappa_{<}} I_{m}[\kappa_{<}]
    \end{pmatrix}.
\label{T-mrep}
\end{eqnarray}
To obtain the above result we have used an identity relating the exponentiated cosine to a sum of modified Bessel functions of integer order
\cite{Abramowitz:70}:
\begin{equation}
\label{Bessel-identity}
 e^{J \cos(\theta)} =
\sum^{\infty}_{m=-\infty} I_{m}(J) e^{i m \theta}.
\end{equation}
The action of the transfer matrix or imaginary-time evolution operator can be further simplified by 
diagonalizing it in the remaining $2 \times2$ subspace of
secondary structure.

One may note that the above matrix (Eq.~\ref{T-mrep})  is
non-Hermitian reflecting the lack of time-reversal symmetry of the underlying Hamiltonian. This absence of time-reversal symmetry occurs because the local bending modulus between the $i^{\rm th}$ and $(i+1)^{\rm th}$ chain tangents depends only on $s_i$, the secondary structure variable to the {\em right} of the first tangent vector. The absence of microscopic left/right symmetry along the chain results in the non-Hermitian character of Eq.~\ref{T-mrep}. Minor modifications of
Eq.~\ref{HCWLC-hamiltonian} generate Hermitian Hamiltonians
expressing the same physics at length scales larger than the monomer size, but we do not pursue such related problems here.

The transfer matrix Eq.~\ref{T-mrep} can be diagonalized in the
space of secondary structure by a similarity transform using the matrix $U(m)$ (defined in appendix \ref{U-mat}); the eigenvalues of the transfer matrix are
\begin{equation}
\label{eigenvalues}
 \lambda_{1,2}(m) =
\frac{\omega_{m}(\kappa_{>})}{2}[ 1 + z_{m} \pm \sqrt{(1-z_{m})^2
+ \beta^2 z_{m}} ],
\end{equation}
where $\omega_{m}(\kappa) = \exp(-\kappa) I_{m}(\kappa)$ is the
transfer matrix element for an ordinary WLC and $z_{m} = e^{-h}
\frac{\omega_{m}(\kappa_{<})}{\omega_{m}(\kappa_{>})}$ is the
ratio of the fugacities of a random coil segment and an
$\alpha$-helical segment for a given angular momentum $m$. The
quantity $\beta$ appearing in the above expression is the
exponentiated free energy cost of introducing a domain wall in the secondary structure, helix-coil variables. $\beta = \exp[\log[(2)- \epsilon_{\rm w}]$ with the first and second terms arising from respectively the entropic gain and enthalpic cost of the creation of a domain wall.

It is important to note that the partition function of the chain
may not be reduced simply to the product of eigenvalues since
doing so presupposes periodic boundary conditions. While in the
thermodynamic limit of long chains ($N\longrightarrow \infty$) it is indeed permissible to choose those nonphysical boundary
conditions, if we wish to study the dependence of mechanical
properties on the degree of polymerization it is essential that we avoid this simplification. Internal consistency
requires that we not impose such boundary conditions on the
secondary structure variables since we cannot impose them on the
chain tangents in the bent configuration. Additionally, such periodic boundary conditions will substantially and artificially reduce the statistical weight of
the appearance of random coil ($s = -1$) segments along the chain. The cause of this artificial reduction is the following. With periodic boundary conditions domain walls must appear in pairs while in the physical problem they can appear individually by destroying the native secondary structure from the ends of the chain. Periodic boundary conditions therefore suppress the probability of the creation of random-coil segments by one extra factor $\beta$ ($\ll 1$ in highly cooperative chains).

The partition function is thus reduced to the remaining sum given
by
\begin{equation}
\label{Z-finalsum} Z = \sum_{s_{0},s_{N},m} \langle s_0 | U(m)
D^N(m) U^{-1}(m) |s_{N} \rangle \eta_{s_{N}},
\end{equation}
where $D(m) = U^{-1}(m)  T(m)  U(m)$ is the diagonalized transfer matrix and the remaining sums are over the secondary structure of the first and last monomers of the chain as well as a single remaining angular momentum variable, $m$. We evaluate
this final sums over $s_0,s_N$ to find:
\begin{eqnarray}
Z(\psi) = \sum^{\infty}_{m= -\infty} e^{i \psi m} [ \frac{1 +
e^{-h}}{2} (\lambda^{N}_{1}(m) + \lambda^{N}_{2}(m)) + \nonumber
\\ \frac{\lambda^{N}_{1}(m) - \lambda^{N}_{2}(m)}{2
\sqrt{(1-z_{m})^2 + \beta^2 z_{m}}}\cdot \nonumber \\ ( 1 +
\beta z_{m} - z_{m} - e^{-h} [ 1 - z_{m} - \beta]) ].
\label{Z-bend-final}
\end{eqnarray}

\begin{figure}
\includegraphics[width=8cm]{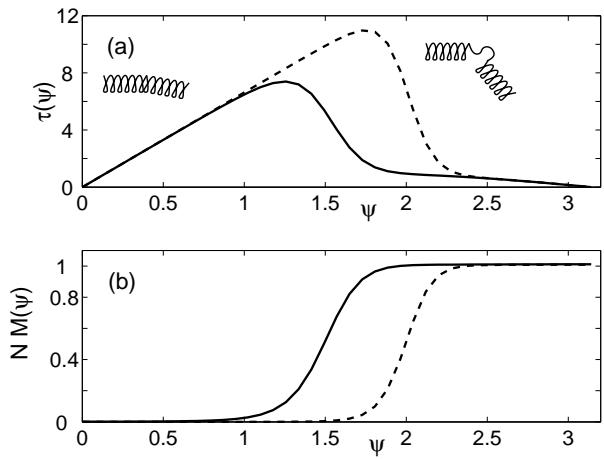}
\caption{\label{torque-curves} a) The torque $ \tau(\psi) $ required to hold the chain at a fixed angle $\psi$ for $\epsilon_{\rm w}=1.4$ (solid line) and $\epsilon_{\rm w}=5.2$ (dashed line) with $h=8$. In both cases $\kappa_{>}=100$, $\kappa_{<}=1$ and $N=15$.  b) The corresponding number of segments having nonnative (random coil) structure $ N  M(\psi)$. The buckling of the alpha-helix coincides with creation of one random coil segment that acts as a softer joint along the very rigid chain. }
\end{figure}

Using the above partition function we may immediately compute two
measurable quantities: (i) the mean torque $\tau(\psi)$ required
to enforce the constraint on the chain tangents at either end, and
(ii) the fraction of molecule in its non-native (random-coil)
structure $M(\psi)$. The former can be directly measured by AFM in
single molecule manipulation experiments while the later can be
probed by circular dichroism spectroscopy.  These quantities can
be computed from the partition function shown in
Eq.~\ref{Z-bend-final} by derivatives
\begin{equation}
\tau(\psi) = \frac{\partial ln Z(\psi)}{\partial \psi}
\label{torque}
\end{equation}
and
\begin{equation}
M(\psi) = - \frac{1}{N} \frac{\partial ln Z(\psi)}{\partial h}.
\label{magnetization}
\end{equation}
Exploiting the symmetry $\omega_m(\kappa) = \omega_{-m}(\kappa)$
we may collect the terms within the $[\cdot ]$ in
Eq.~\ref{Z-bend-final} defining $ [\cdot] = \mathcal{Z} (m,h) $ to
write the partition function as
\begin{equation}
Z(\psi) = \mathcal{Z}(0, h) + 2 \sum^{\infty}_{m=1} \cos(\psi m)
\mathcal{Z}(m, h) \label{parfnhalfmsumecfexp}
\end{equation}
so that, using Eqs.~\ref{torque},\ref{magnetization} we find
\begin{equation}
\tau(\psi) = \frac{ - 2 \sum^{\infty}_{m=1} m \sin(\psi m)
\mathcal{Z}(m, h)}{\mathcal{Z}(0, h) + 2 \sum^{\infty}_{m=1}
\cos(\psi m) \mathcal{Z}(m, h)} \label{torque-sum}
\end{equation}
and
\begin{equation}
M(\psi) = \frac{-1}{N} \frac{ \frac{\partial \mathcal{Z}(0, h)}{\partial h} + 2 \sum^{\infty}_{m=1} \cos(\psi m) \frac{\partial \mathcal{Z}(m,
h)}{\partial h}}{\mathcal{Z}(0, h) + 2 \sum^{\infty}_{m=1}
\cos(\psi m) \mathcal{Z}(m, h)} \label{mag-sum}.
\end{equation}
The series above are well approximated by partial sums; in practice, taking the first ten terms reduces the error to about one part in $10^{6}$. These numerically evaluated partial sums are shown in figure \ref{torque-curves}.  At small values of the bending angle $\psi$, there is a linear dependence of the constraining torque on $\psi$. The alpha-helix bends like a flexible, elastic rod. At a certain critical angle $\psi^\star$, however, the constraining torque reaches a maximum and then drops precipitously for angles $\psi > \psi^\star$  as shown in part (a) of figure~\ref{torque-curves}. This dramatic collapse of the chain's rigidity is akin to the buckling instability of a macroscopic tube such as a drinking straw. The mode of the localized failure is, however, completely different. Examining part(b) of figure \ref{torque-curves}, we see that at $\psi = \psi^\star$, $M$ abruptly jumps to  $\mathcal{O}(1/N)$. The buckling of the alpha-helix is due to the creation of a single random coil segment along the chain that provides a region of greatly reduced bending stiffness. The size of the created random-coil section  will remain on the order of $N \kappa_</\kappa_>$ so for a large difference in bending moduli between the native and nonnative states of the chain, these ``weak links'' generically occupy a small fraction of the polymer. In the above example, there is only one weak link.

The underlying cause of the buckling can be understood in terms of a comparison of the free energy increase per monomer associated with the creation of nonnative states along the chain due to the reduction in chain bending energy due to the collapse of chain curvature into the more flexible random coil segment. At  large enough imposed curvatures (at $\psi^\star$) the reduction of chain bending energy more than offsets the free energy cost associated with the creation of a chain segment having nonnative secondary structure and the buckling of the alpha-helix occurs. The mean conformation of chain changes from the arc of a circle (to minimize curvature energy) to two essentially straight segments connected by $\mathcal{O}(1)$ monomers in the random coil state where the externally imposed curvature localizes.  It should be noted that critical angle $\psi^\star$ for the buckling of an alpha-helix depends not only on relative persistence lengths in the native and nonnative states of the chain, but also on the helix/coil parameters. If, for example, the chain cooperativity is decreased by changing $\epsilon_{\rm w} = 5.2 \longrightarrow 1.4$ (dashed line to solid line in figure~\ref{torque-curves}) then the free energy cost of creating a weak joint in the chain is reduced and the buckling transition occurs at smaller imposed curvatures. One may observe similar effects through the change in $h$. 

The buckling transition may be considered as a type of nucleation process of random coil segments along the chain. Using this analogy, we can distinguish three different types of buckling transitions based on the length of the chain and the chain cooperativity. For very long chains there is homogeneous nucleation in which the random coil segment first appears anywhere in the bulk of the chain, {\em i.e.} away from the chain ends. The creation of a random coil segment in the bulk of the chain requires the creation of two domain walls whereas the creation of random coil segments at the end of the chain requires only one. For long enough chains the added enthalpic cost of the additional domain wall is more than compensated by the translational entropy associated with the placement of that random coil segment. For shorter chains or chains having higher cooperativity, homogeneous nucleation is replaced by heterogenous nucleation; the random coil segment appears at one end of the chain. Based on these considerations, the transition between homogeneous and heterogeneous nucleation of random coil segments should occur for chains where $\log N \sim \epsilon_{\rm w}$. Finally, even shorter chains or chains with still higher chain cooperativity, one can encounter a regime in which the entire chain spontaneously loses its native secondary structure at a critical angle. Such a transition should occur only if $ N < \epsilon_{\rm w}/h$.   The curves shown in figure~\ref{torque-curves} correspond the intermediate case of heterogeneous nucleation of random coil segments. 

The appearance of the buckling transition is the first, and perhaps most dramatic consequence of the coupling the helix/coil, internal state variables of the chain to its conformational degrees of freedom. The generic consequence of such a coupling is the highly nonlinear bending elasticity of the polymer as shown in figure~\ref{torque-curves}. We now consider the effects secondary structure on the force extension relations of the polymer.

\subsection{Stretching}
\label{stretching}

Before we develop the theory of stretching the HCWLC, we discuss the radius of gyration of such a polymer \cite{Debye:27,Benoit:53}. To do so we note that the distance between the $i^{\rm th}$ and $j^{\rm th}$ monomers along the chain is given by
\begin{equation}
\label{Rij}
R_{ij} = \sum^{j-1}_{n=i} \gamma (s_n) \hat{t}_{n},
\end{equation}
where $\gamma (s_n)$ is the length of a monomer measured along its mean chain tangent. Recalling that each monomer of the HCWLC represents enough amino acids ($\sim 3$) to unambiguously assign a secondary structure to the segment, the effective size of the monomer depends on that secondary structure. In the native, alpha helix state the monomer is shorter than in its generally more extended random coil configuration. To account for this aspect of the coarse-grained polymer model we define a monomer length that is a function of the secondary structure variable $s_n$ via
\begin{equation}
\label{gamma-def} \gamma(s) = \left\{ \begin{array}{ll}
                        \gamma_< & \mbox{if $s = + 1$} \\
                        \gamma_> & \mbox{if $s=-1$}
                   \end{array}
            \right.  ,
\end{equation}
where, as the notation suggests, $\gamma_< < \gamma_>$, the length of a monomer increases when it loses its alpha-helical secondary structure. We return to a discussion of reasonable numerical estimates of these values in the conclusions below.  

\subsubsection{The radius of gyration}

\begin{figure}
\includegraphics[width=8cm]{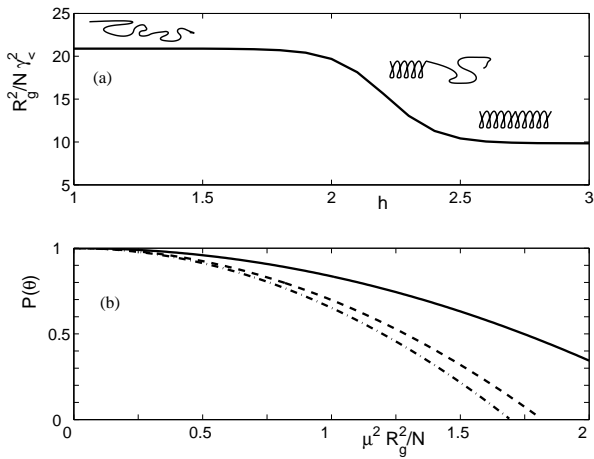}
\caption{\label{scattering-figure} a) The radius of gyration of the HCWLC as a function of the free 
energy cost per segment to transform to the random coil, nonnative state: $h$. In this curve 
$\kappa_{>} = 100$, $\kappa_{<} = 1$, $N = 10$, and $\epsilon_{\rm w} = 10$. In the lower figure b) we 
plot the scattering function $P(\theta)$ computed from the radius of gyration. The dashed line 
corresponds to that of a chain that is a mixture of helix and coils for $h=2.1$ and 
$\epsilon_{\rm w} = 10$. The solid line and dashed-dotted lines are the scattering functions 
computed for the molecule in the all helix and all random-coil states respectively.} 
\end{figure}

We now compute the radius of gyration of the these polymers in solution by taking the thermal average of $R_{0N}^2$ using Eq.~\ref{Rij} and the HCWLC Hamiltonian. Separating the sum into terms diagonal and off-diagonal in the monomer indices we write
\begin{equation}
\label{radius-gyration}
\langle R^{2}_{0N} \rangle  = \sum^{N-1}_{n=0} \langle \gamma^{2}(s_n) \rangle +
2 \sum^{N-1}_{n=0} \sum^{n}_{m=1} \langle \gamma(s_n) \gamma(s_m) \hat{t}_n \cdot \hat{t}_{m} \rangle.
\end{equation}
Both of these terms in the above expression can be evaluated in terms of the transfer matrices introduced in Eq.~\ref{T-mrep}. We explore this point in some detail as it gives insight into the bulk of the calculations regarding stretching of the HCWLC. The reader who is uninterested in details of the calculation can pick up the discussion involving figure \ref{scattering-figure} in the last two paragraphs of this section. 

We note that the first (diagonal) term may be written in terms of the transfer matrices as
\begin{eqnarray}
\langle \gamma^{2}(s_k) \rangle =  \sum_{s_0, s_N,m} 
\frac{\eta_{s_N}}{4 \pi^2 Z}  \nonumber \\
  \int_{\theta_0,\theta_N} e^{i m (\theta_N - \theta_0)}   \langle s_0,m | D^k(m) \Gamma^2 D^{N-k}(m) | s_N,m  \rangle, 
\label{diagonal-term-length}
\end{eqnarray}
where both integrals range over the full unit circle.
In Eq.~\ref{diagonal-term-length} $Z$ is the partition function defined in Eq.~\ref{Z-quantum} and we have also introduced the matrix $\Gamma$ defined by
\begin{equation}
\label{Gamma-matrix}
\Gamma =
    \begin{pmatrix}
     \gamma_{<}      &  0 \\
     0        & \gamma_{>}
    \end{pmatrix}.
\end{equation}
This matrix acting in space of secondary structure assigns the appropriate monomer length to the segment, {\em i.e.} $\langle +1,m | \Gamma | \ +1, m'\rangle = \delta_{m,m'} \gamma_<$ and $\langle -1,m | \Gamma| \ -1, m'\rangle = \delta_{m,m'} \gamma_>$ while both off-diagonal terms vanish. It represents the action of the $\gamma(s)$ operator acting on a given state of the HCWLC, $| s,m \rangle$. The transition amplitude appearing in the above equation may be interpreted again as the amplitude for the fictitious quantum particle. In this case we compute the amplitude to propagate $k$ imaginary time slices at angular momentum $m$, be acted on by $\Gamma^2$ that measures the square of the length of $k^{\rm th}$ segment and then propagate the remaining $N-k$ imaginary time steps at the same angular momentum.   Finally, the integrals over the initial and final angles of the chain may be performed explicitly; by not constraining these two end tangents we project out the $m=0$ state of the chain so that Eq.~\ref{diagonal-term-length} may be simplified to
\begin{equation}
\label{diagonal-term-length-II}
\langle \gamma^{2}(s_k) \rangle =  \sum_{s_0, s_N} 
\frac{\eta_{s_N}}{ Z}  \langle s_0,m=0 | D(0)^k \Gamma^2 D(0)^{N-k} | s_N,m=0  \rangle.
\end{equation}
The remaining sums over the secondary structure of the initial and final chain monomers consists of only four terms and can be evaluated directly.

Using the similar reasoning we may write the off-diagonal parts of Eq.~\ref{radius-gyration} as
\begin{eqnarray}
\langle \gamma(s_k) \gamma(s_j) \hat{t}_k \cdot \hat{t}_j \rangle =  \nonumber \\
 2 \sum_{s_0, s_N} 
\frac{\eta_{s_N}}{ Z}  \langle s_0,0 | D(0)^k \Gamma D(1)^{j-k} \Gamma D(0)^{N-j} | s_N,0  \rangle,
\label{offdiagonal-term-length}
\end{eqnarray}
where we have taken $j>k$.  We observe that while the initial and final states are fixed at zero angular momentum (for the same reasons as in Eq.~\ref{diagonal-term-length-II}), the diagonalized transition matrix acting between the $k^{\rm th}$ and $j^{\rm th}$ monomers is evaluated at an angular momentum of unity.  To understand this we note that the thermal average of the scalar product in the above equation involves the averages of products of cosines of the form $\cos(\theta_j) \cos(\theta_k)$. These cosines generate $m \longrightarrow m \pm 1 $ transitions in the angular momentum basis so that the action of the cosine at the $k^{\rm th}$  monomer takes the initial $m=0$ state into either of $m=\pm 1$ states. The action of second cosine at the $j^{\rm th}$  monomer must return the angular momentum of the state to zero so that the integral over the initial and final angles does not cause this contribution to the transition amplitude to vanish. Thus of the four possible combination of $m \longrightarrow m \pm 1 $ acting at the two sites, only the two terms leading to no net change in angular momentum survive the final averaging. Because $D(m)$ is even in $m$, the factor of two accounts for both of these terms. Once again the remaining sums over the secondary structure of the initial and final chain segments can be performed directly.

The final sums over $j,k$ required to determine the radius of gyration can be performed. If one were to assume translational invariance along the chain, the remaining sums over monomers in Eq.~\ref{radius-gyration} can be rewritten as a single sum
\begin{equation}
\label{single-sum}
\langle R^{2}_{0N} \rangle  = \sum^{N-1}_{k=0} (N-k) C(k),
\end{equation}
where we have defined the quantity $C(k) = \langle \gamma(s_{i+k}) \gamma(s_i) \hat{t}_{i+k} \cdot \hat{t}_i \rangle$ to be the tangent vector correlation function function weighted by the length of the chain segments. Due to our momentary assumption of translational invariance along the chain, this function is independent of the monomer index $i$.  With this assumption the correlation function simplifies to 
\begin{equation}
\label{correlation-function}
C(k) = \frac{1}{Z} \mbox{Tr} \left[ \Gamma^{(01)} \cdot D(1)^k \cdot \Gamma^{(10)} \cdot D(0)^{N-k} \right],
\end{equation}
where we have introduced the matrices: 
\begin{eqnarray}
\label{Gamma-01}
\Gamma^{(01)} &=& U(0)^{-1} \cdot \Gamma \cdot U(1)\\
\Gamma^{(10)} &=& U(1)^{-1} \cdot \Gamma \cdot U(0)
\label{Gamma-10}
\end{eqnarray}
written in terms of $\Gamma$ defined in Eq.~\ref{gamma-def} and the matrices $U(m)$ that diagonalize the transfer matrix at angular momentum $m$ (see appendix \ref{U-mat}). By performing the above trace,  
going to the limit of large $N$, and using fact that at any given angular momentum the helix phase is more probable than the random coil ({\em i.e.\/} $\lambda_1(m) > \lambda_2(m)$ for all $m$),   we may write the correlation function as the sum of two exponentials:
\begin{equation}
\label{correlation-function-II}
C(k) = A e^{-k \ell_a} + B e^{-k \ell_b},
\end{equation}
where the two lengths controlling the exponential decay of correlations are given in terms of the eigenvalues (Eq.~\ref{eigenvalues}) by
\begin{eqnarray}
\ell_a &=& \log \left( \frac{\lambda_1(0)}{\lambda_1(1)} \right) \\
\ell_b &=& \log \left( \frac{\lambda_1(0)}{\lambda_2(1)} \right).
\end{eqnarray}
The coefficients of these two decays are written explicitly in the appendix \ref{appendix}. 

The existence of two exponential decay lengths for the correlation function reflects the fact that between any two tangents along the chain, the polymer may exist in one of two states having differing monomer lengths and thermal persistence lengths. To better understand this result it is instructive to study the limit in which it is highly unlikely to find the chain in the random coil state: $\epsilon_{\rm w} \gg 1$ and $h \gg 1$. In that limit, $\lambda_2(m) \longrightarrow 0$ so that $\ell_b \longrightarrow \infty$; the decay of correlations is dominated by the length $\ell_a \sim 1/\kappa_>$ in this limit. From extensions of this reasoning one can associate $\ell_a$ with the decay length of  correlations for a section of polymer that starts and ends with helical segments. Between these helical segments this correlation length is related to some function of {\em both\/} $\kappa_<,\kappa_>$ due to fluctuations into the coil phase.  Similarly, the length $\ell_b$ controls the decay of correlations for segments of the chain that begin and end with random coil segments. Once again, due to fluctuations into the helical phase this length is a function of both $\kappa_<,\kappa_>$.  

Finally, given a form for the correlation function $C(k)$ obtained from the assumption of translational invariance along the chain, we may directly evaluate the radius of gyration. We find:
\begin{eqnarray}
\langle R^{2}_{0N} \rangle  = \frac{A e^{\ell_a}}{(e^{\ell_a}-1)^2} \left[ N (e^{\ell_a}-1) + e^{-N \ell_a} -1 \right] + \nonumber \\
+ \frac{B e^{\ell_b}}{(e^{\ell_b}-1)^2} \left[ N (e^{\ell_b}-1) + e^{-N \ell_b} -1 \right].
\label{single-sum-answer}
\end{eqnarray}

We suspect, however, that at least in the case of highly cooperative or short chains for which $\epsilon _{\rm w} \gg \log N$ the effects of the chain ends will be significant and thus break the assumed translational invariance used above. In order to evaluate the radius of gyration of the chains that strongly break translational invariance, we numerically evaluate the requisite sums to determine the radius of gyration; these results are shown in figure \ref{scattering-figure}. There we see the cross-over of a random coil ($N > \kappa_<$) having a Kuhn length of $\gamma_>=3.0$ to an essentially straight rod of length $\gamma_< N$ in the helix phase.  

The radius of gyration can be experimentally  probed via small angle elastic scattering. In figure \ref{scattering-figure} (b) we plot the scattering function for HCWLCs in dilute solution as a function of the scattering wavevector for scattering angle $\zeta$: $\mu = \frac{4 \pi}{\lambda} \sin(\zeta/2)$ In the same figure we show the predicted scattering from WLCs. 

A comparison of the predicted scattering HCWLCs and the better studied WLCs demonstrates that such scattering experiments alone are ineffective in differentiating between these two models. The scattering from a HCWLC can always be interpreted in terms of the scattering form a WLC having some effective persistence length. More generally, any measure of the radius of gyration will not distinguish the HCWLC from a simple WLC as long as the effective persistence length of the chain is adjusted to fit the data. In order to observe qualitatively novel behavior of the HCWLC, one must probe the force extension behavior of the chains. Here we will see highly nonlinear elasticity mirroring the nonlinear bending elasticity of these polymers.

\subsubsection{Force extension relations: Small forces}

In the presence of a stretching force $F$ the Hamiltonian of the HCWLC may be written as
\begin{equation}
H = H_{0} - F \sum^{N}_{i=0} \gamma(s_{i}) \cos(\theta_{i}),
\label{Hamiltonian-stretching}
\end{equation}
where $H_{0}$ represents the HCWLC Hamiltonian in the absence of externally applied forces as shown in Eq.~\ref{HCWLC-hamiltonian}. A calculation of the chain partition function based on the above Hamiltonian would, of course, result in the complete description of the equilibrium force/extension relations for this model. Unfortunately, a closed form expression for this partition function is not possible since $H_0$ and the term proportional to the applied force are diagonalizable in the momentum and position representations respectively. The basis states that diagonalize the full Hamiltonian, Eq.\ref{Hamiltonian-stretching}, are the energy eigenstates of the quantum pendulum.  We do not pursue this approach here.  The identical issue arose for the study of the stretched WLC; there approximate numerical diagonalization \cite{Marko:95} and variational calculations \cite{Marko:95,Lamura:01,Storm:03} have been successfully employed. 

We begin by considering small externally applied forces  and consequently small chain extensions $\Delta L$. Defining $\Delta \gamma = \gamma_> - \gamma_<$ to be the extension of a monomer under the helix-to-coil transition, we consider the small force to be those for which $F \Delta \gamma$ is small in comparison to the other four energy scales in the problem. Using this assumption, it becomes reasonable to expand the chain free energy in powers of the externally applied force. We thus generate a cumulant expansion
\begin{equation}
\label{cumulant-expansion}
\log Z(\psi) = \sum^{\infty}_{l=0} \frac{c_{l}}{l!} F^{l}
\end{equation}
that is similar in spirit to those obtained from high temperature expansions of the Ising model \cite{Domb:74}. In the above equation $c_l$ is the $l^{\rm th}$ order cumulant. These cumulants are thermally averaged quantities in which the averaging is performed with respect to the zero applied force Hamiltonian, $H_0$. Calculating derivatives of Eq.~\ref{cumulant-expansion} evaluated at $F=0$ allows one to calculate the mean extension of the chain in the direction of the applied force in powers of $F$.  Finally, we note that since the remaining thermal averages are to be performed with respect to $H_0$, we may borrow the formalism used to compute the torque/angle curves and consider averages over restricted ensembles in which the first chain tangent is directed along $\vec{F}$ (taken to be in the $\hat{x}$-direction) while  the last chain tangent is fixed at an angle $\psi$. We can thereby explore the coupling of applied torques to the extensional compliance of the chain using this formalism.

Taking the first cumulant, which is the term linear in the force, we calculate the mean length of the polymer  chain in the {\em absence} of any applied force
\begin{equation}
\label{Mean-Length-smallF}
\langle L \rangle (\psi) = \langle \sum^{N}_{k=0} \gamma(s_k) \cos \theta_k \rangle_\psi,
\end{equation}
where the average $\langle \cdot \rangle_\psi $ is taken over the force-free, restricted ensemble of chains having an initial tangent in the $\hat{x}$ direction and a final tangent making an angle $\psi$ with respect to that initial tangent.  The restriction placed on the first tangent breaks the rotational symmetry of the system leading to a nonvanishing value of $\langle L \rangle (\psi)$.  

\begin{figure}
\includegraphics[width=6cm]{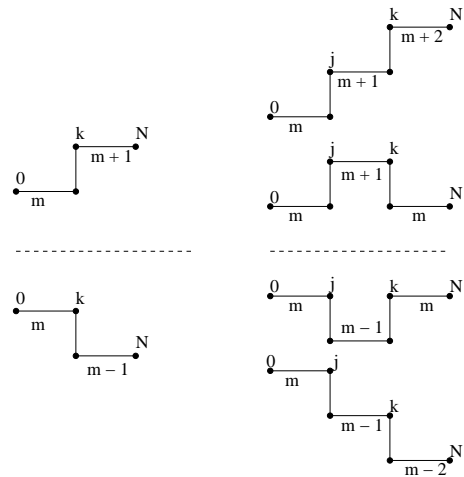}
\caption{\label{p-theory} Diagrammatic expansion used to calculate the extension of the chain order-by-order in the externally applied force.  The combination of two terms on the left give the \protect{$\mathcal{O}(F)$} terms that determine the mean length of the chain at zero applied force. The four terms on the right give the linear response of the mean length due to an externally applied force. In both figures the angular momentum increases in the vertical direction and each horizontal leg of the walks represent products of the transfer matrices at the labelled angular momentum. The labels $0$, and $N$ denote the beginning and end of each chain respectively. The labelled intermediate sites $k$ and $j$ denote the monomers where the cosines act to either increase of decrease the angular momentum of the walk by one unit. Mirror reflections of each diagram about the dotted line has the same contribution
to the sum.}
\end{figure}

Based on our discussion of the radius of gyration of the chain, we can compute these mean extension at zero force using the transfer matrix technique; Eq.~\ref{Mean-Length-smallF} may be written as
\begin{equation}
\langle L \rangle = 
 \sum^{N}_{k=0} \sum_{s_0, s_N} \langle s_0, \theta_0=0 \vert T^k \gamma(s_k) \cos \theta_k  T^{N-k} \vert s_N, \psi \rangle \eta_{s_{N}}.
\label{Mean-L-II}
\end{equation}
Working, once again, in the momentum representation we can recast the above expression into a simple sum over one angular momentum variable and the four possible combinations of secondary structure states of the initial and final chain segments:
\begin{eqnarray}
\langle L \rangle(\psi) = \sum^{N-1}_{k=0} \sum_{s_{0}, s_{N}}  \eta_{s_{N}}[ \langle s_{0},0 | T^{k} \Gamma T^{N-k} | s_{N},1 \rangle \cos(\psi) + \nonumber \\ 2 \sum^{\infty}_{m=1} \langle s_{0},m | T^{k} \Gamma T^{N-k}| s_{N},m+1 \rangle \cos([m+1] \psi)]. \nonumber \\
& & 
\label{Mean-L-m}
\end{eqnarray}
The structure of the above expression may be characterized by using a simple graphical representation. In the left panel of figure \ref{p-theory}, we represent the above terms for the first cumulant (the mean length of the chain) as the set of all one-step random walks in angular momentum space. As noted above in the computation of the radius of gyration, each factor of cosine increments or decrements the angular momentum.  In general the $n^{\rm th}$-order cumulant requires the determination of the thermal average of the product of $n$ such cosines along the chain contour: $ \langle \cos \theta_{i_1}  \cdot \cos \theta_{i_2} \cdots \cos \theta_{i_n} \rangle$. There is a one-to-one mapping of such products to the set of all $n$-step random walks in momentum space. For example, the right hand panel of figure \ref{p-theory} shows all two-step random walks. That set of random walks in the momentum space sum to give the linear response of the mean length of the chain to the externally applied force $F$. 

By expanding about zero applied force we may obtain an expression for the series expansion of the mean length as a function of applied force of the form
\begin{equation}
\label{Length-expansion}
\langle L \rangle(F,\psi) = L_0(\psi) +  L_{1}(\psi)  F + L_{2}(\psi) F^2 + \ldots, 
\end{equation}
where the prefactors of the odd-index terms, {\em i.e.} $L_{2n+1}(\psi) F^{2 n +1}$ vanish upon averaging over all end angles, $\psi$. Each term in the above expansion is the set of all $n$-step random walks in momentum space. Each involves a sum over the states of secondary structure at each end of the chain, which may be performed exactly and one infinite over a single angular momentum variable. The latter sum cannot in general be performed exactly, but, as discussed above, it may be numerically approximated to arbitrary precision. In practice because of the rapid convergence of this sum with high $m$, only a few terms are required to generate an excellent approximation.

Constraining the initial and final chain tangents causes each walk in momentum space to be weighted by a phase factor $\exp (i \Delta m \, \psi)$. Thermal averages over the ensemble of chains having unconstrained final tangents can be computed by averaging over $\psi$.  Due to the aforementioned $\psi$-dependent phase factor, this averaging eliminates all walks that results in a net change in the angular momentum of the chain. For instance the two walks comprising the mean length of the chain at zero force (left panel of figure \ref{p-theory}) both vanish when the final angle is unconstrained. This is to be expected from basic symmetry considerations. By relaxing  that constraint, one restores the rotational symmetry of the problem so that the mean extension of the polymer along the $\hat{x}$ axis necessarily vanishes in the limit of zero applied force.  

Since the transfer matrix is even with respect to angular momentum, there is an additional reflection symmetry; each walk from $m$ to $m'$ makes an identical contribution to the final result as that walk reflected about $m=0$, {\em i.e.} the walk from $-m$ to $-m'$ in which each increment of angular momentum is replaced by a decrement and vis versa. We employ this additional symmetry of the problem to rewrite the set of two random walks comprising the mean length as shown in Eq.~\ref{Mean-L-m}.  

\begin{figure}
\includegraphics[width=6cm]{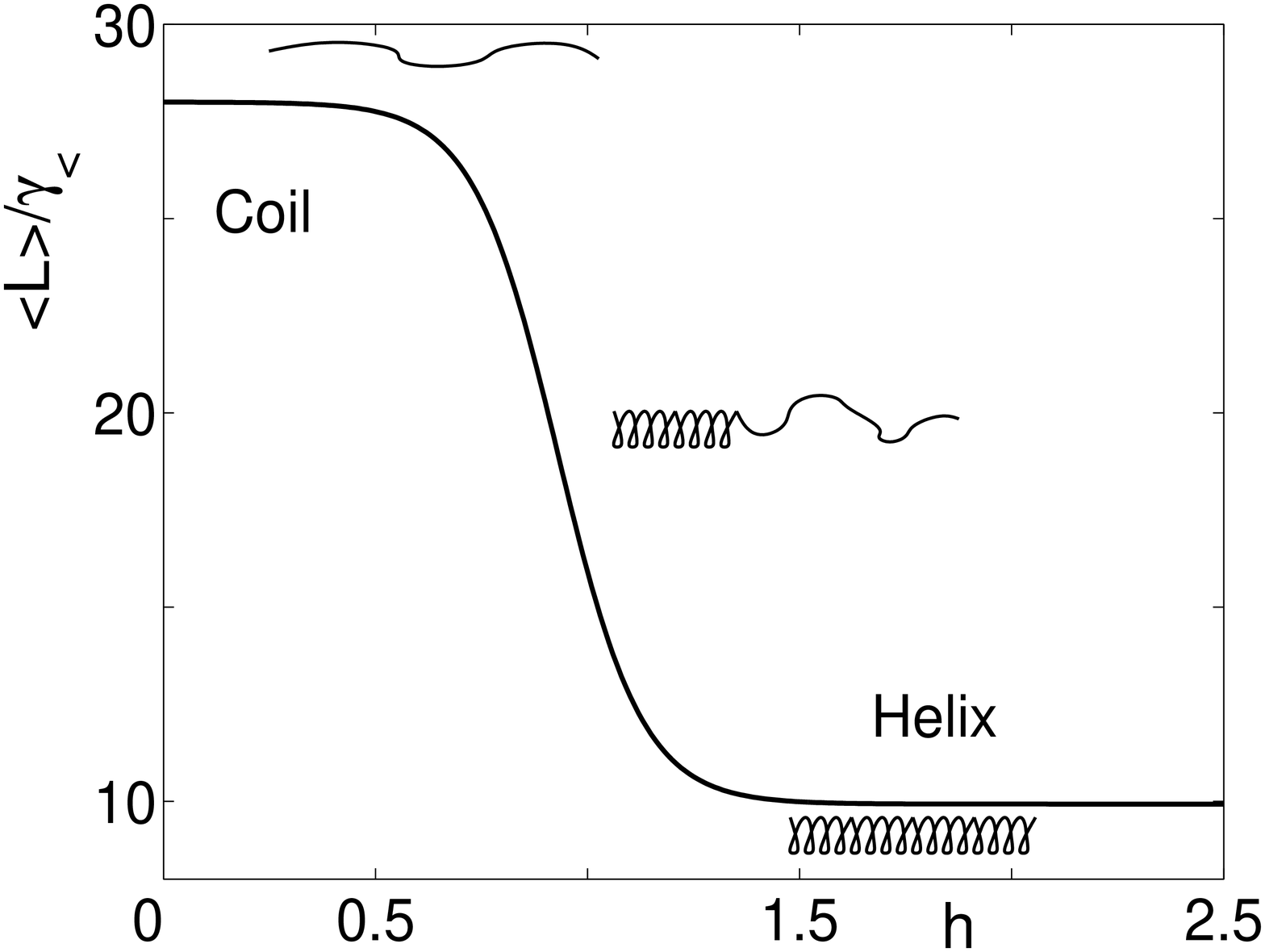}
\caption{\label{mean-length-fig} 
Mean length of the HCWLC in thermal equilibrium as a
function of the parameter $h$ -- the free energy cost per segment to transform to the random coil, nonnative state. The bending modulus of the coil section $\kappa_{<}=10$, $\kappa_> = 100.$ and the length of the helix section $\gamma_{<}=1$ for this plot. The polymerization index of the chain is $N = 10$. The initial and final chain tangents are fixed so that $\theta_0 = \theta_N = 0$.  As $h$ increases we see the effect of the HC transition on the mean length of this stiff chain.}
\end{figure}

We plot in figure \ref{mean-length-fig} the mean length of the chain as a function of $h$, the excess free energy per unit length associated with the existence of nonnative secondary structure. If $h< 1$, we expect the chain to be driven into a random coil, nonnative structure in order to increase the chain conformational entropy associated with the disordering of the polymer backbone tangent vectors. For values of $h \ge 1$ we expect a highly cooperative ($\epsilon_{\rm w} > \log N$ in this example system) and therefore sharp transition to the alpha-helical, native state. This transition is evidenced by the precipitous decrease of the chain's extension occurring at $h \simeq 1$.  

\begin{figure}
\includegraphics[width=6cm]{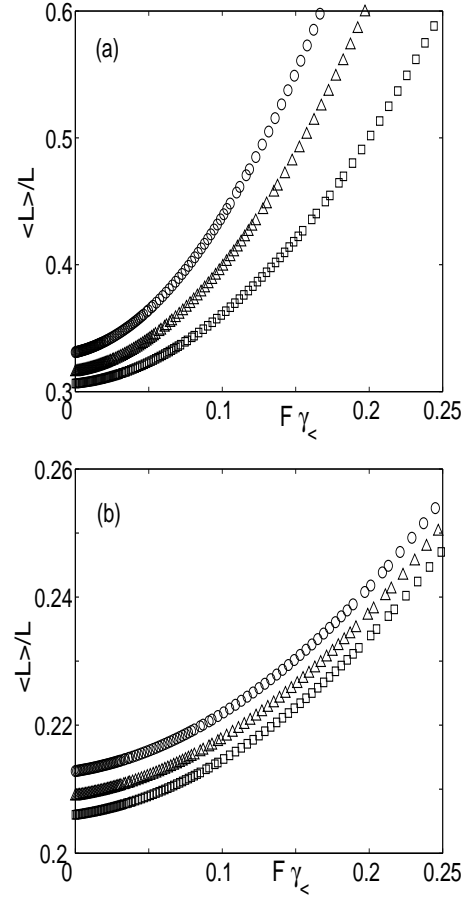}
\caption{\label{fixed-angle-stretching} The mean length vs.\ applied force for HCWLCs with constrained end tangents so that $\theta_0 = 0$ and $\theta_N = \psi$.  In both figures three different final angles are shown: $\psi = 0$ (circles), $\psi = \pi/6$ (triangles) and $\psi = \pi/4$ (squares). In a) $\kappa_{>} = 100,  \kappa_{<} = 1, \epsilon_{\rm w} = 10, h = 3.0, N = 10, \gamma_{>}=3, \gamma_{<}=1$. In b) $\kappa_{>} = 5, \kappa_{<} = 1, \epsilon_{\rm w} = 10, h = 1.5, N = 24$. The upper figure shows the expected behavior of an alpha-helical polypeptide chain. For comparison, a longer HCWLC of shorter persistence lengths is shown in b)}
\end{figure}

Using the formalism described above we may also compute the linear and {\em nonlinear} response of the chain to a force $F$ by considering longer walks in momentum space. We report those results as a function of both $F$ and $\psi$, the angle of final chain tangent. It is likely that in future single molecule force spectroscopy experiments it will be problematic to simultaneously control both the applied force and final chain tangent. While we suspect that force spectroscopy with unconstrained angles will be more experimentally relevant, we believe that in order to discuss alpha-helical domain extensional elasticity within the native state of a protein, such boundary condition prescriptions may prove necessary. We plot in figure \ref{fixed-angle-stretching} the nonlinear force/extension behavior (including terms up to $F^2$) of two representative HCWLCs for a variety of angular bends, $\psi$. Higher order terms in the applied force can be computed similarly. For the case of unconstrained initial and final tangents, we have developed an automated procedure to calculate terms of the perturbation expansion to arbitrary order \cite{Chakrabarti:04}.

In figure \ref{fixed-angle-stretching}(a) we observe the nonlinear extensional compliance (through terms of order $F^2$) of the HCWLC for the case that chain is much shorter than its thermal persistence length. The different symbols correspond to differing imposed curvatures on the chain -- see the caption.  The finite extension at zero force is consistent with the assumption that the chains are simply bent into the arc of a circle consistent with uniform distribution of chain curvature as required to minimize chain bending energy. By noting the difference in slopes of the three extension vs.\ force curves in figure \ref{fixed-angle-stretching}(a), one sees that the extensional compliance of the chain in the direction parallel to the applied force depends on the angle of the final chain tangent, $\psi$.  In part (b) of this figure we observe the predicted force extension relations  to the same order in the low-force perturbation theory for the HCWLCs where the persistence length is less than $L$ even in the stiffer, alpha-helical phase.  As evidenced by the equality of the slopes of all three figures, the effect of imposed curvature on the subsequent extensional compliance of this chain is minimal. For a more flexible chain twisting the final tangent has little effect on the compliance under subsequent extension. We expect the upper panel of figure  \ref{fixed-angle-stretching} to better reflect the mechanical of alpha-helical polypeptides.

The perturbation theory presented above appears to be a useful approach to the study of small extensional deformations of alpha-helical protein domains. A generic feature of this model is the nonlinear growth of the effective extensional compliance of the chain with increasing applied force. The underlying mechanism must be the enhanced statistical weight for finding a segment in its {\rm longer} nonnative state. The applied force thereby accesses a reservoir of chain length built into each segment; such a mechanism has been studied in detail by Tamashiro and Pincus \cite{Tamashiro:01} although that work ignored the role of the tangent vector fluctuations. These fluctuations should only enhance the nonlinear growth of the effective chain compliance. When $\kappa_<  \ll \kappa_>$ the random coil sections of the chain recover a significant entropic contribution to their free energy. The chain as it extends into the random coil phase not only becomes physically longer, but also has a decreasing effective persistence length. The combination of these two factors enhance the effective compliance of the chain.

The perturbatively evaluated HCWLC model, however, fails to reproduce the central aspect of the high force limit. In order to capture this basic feature of the WLC that must also hold for the HCWLC, we must augment our low-force perturbation theory by other methods better adapted to the high force limit.  Since one may compute the  $n^{\rm th} $-cumulant, which generates the $(n-1)^{\rm th}$ term of the series expansion of the force extension curve as shown in Eq.~\ref{Length-expansion}, one might at first imagine that one could perturbatively determine the force extension relation of the HCWLC at arbitrarily large forces. This is not the case. In fact, the perturbative approach to the high force limit is incapable of capturing the essential aspects of the behavior of the WLC  or the more complex HCWLC. The high force limit, $F \longrightarrow \infty$ is an essential singularity of the partition function and thus cannot be approximately by a Laurent series in force. This has been directly confirmed by Marko and Siggia \cite{Marko:95} who have shown by integrating out the transverse contour fluctuations of the WLC in the high force limit that the mean length approaches the maximal length, $L$ as $\langle L \rangle /L \sim 1 - F^{-1/2}$.  Perturbation theory in the low-force limit (expansions in $F$) or in the high force limit (expansions in $1/F$) will miss this result.

\subsubsection{High Force Limit: Mean Field Theory}
\label{Stretch-high-force}

We now study the extensional compliance of the HCWLC in the high force limit. In order to explore the approach under high forces of the mean chain extension to its maximal extension $L = \gamma_> N$, it is reasonable to assume that the tangent vector fluctuations become small so that a Gaussian approximation is justified. We may approximate the HCWLC Hamiltonian Eqs.~\ref{HCWLC-hamiltonian},\ref{Hamiltonian-stretching} by
\begin{eqnarray}
 H = \frac{\epsilon_{\rm w}}{2} \sum^{N-1}_{i=0}
(1 - s_{i} s_{i+1}) - \frac{h}{2} \sum^{N}_{i=0} (s_{i} -1) + \nonumber \\
\sum^{N-1}_{i=0} \frac{\kappa(s_{i})}{2} [\theta_{i+1}-\theta_{i}]^{2} -  
F \sum^{N}_{i=0} \gamma(s_{i}) [ 1 - \frac{\theta^{2}_{i}}{2} ].
\label{HCWLC-Gaussian}
\end{eqnarray} 
The above Hamiltonian is now quadratic in the angles of the chain tangent vectors. We fix the initial and final chain tangents to lie along the direction of the applied force, $\theta_0 = \theta_N = 0$. By iterative Gaussian integrals over the remaining angles we determine an effective partition function that is now a sum over only the secondary structure degrees of freedom. After integrating over $\theta_1, \ldots, \theta_{N-1}$ the partition function in the high force limit reduces to the form
\begin{eqnarray}
Z = \sum_{\left\{s_i\right\}} \exp \left[ \frac{\epsilon_{\rm w}}{2} \sum^{N-1}_{k=0} ( 1 - s_k s_{k+1}) + \right. \nonumber \\
\left. \frac{h}{2} \sum^N_{k=0}(s_k - 1)  
- F \sum^N_{k=0} \gamma(s_k)  \right]
 J \left[\left\{s_i \right\}\right],
 \label{partition-function-high-F}
\end{eqnarray}
where the remaining sum is over all $2^N$ configurations of the secondary structure variables.  We have introduced the quantity $J[\left\{s_i \right\},N]$ produced by the Gaussian integrals. It is a function of that secondary structure configuration defined by
\begin{equation}
\label{High-force-product}
J \left[\left\{s_i \right\}, N\right]=  \prod^N_{j=1} \sqrt{\frac{ 2 \pi}{2 R_j + \kappa(s_j)}} e^{- \gamma(s_0) F}.
\end{equation}
Each term $R_i$ in the above product is defined recursively by the equation
\begin{equation}
\label{R-def}
R_{i} = \frac{\kappa(s_{i-1}) R_{i-1}}{ 2 R_{i-1} + \kappa(s_{i-1}) } + \frac{F \, \gamma(s_{i})}{2}
\end{equation}
for $i = 2,\ldots,N$, where we fix the initial condition for the recursion by setting  
\begin{equation}
\label{R-def-0}
R_{1} = \frac{\kappa(s_0) + F \gamma(s_1)}{2}.
\end{equation}
The $i^{\rm th}$ term in the product depends on the full set of secondary structure variables from site $i-1$ back to $0$. Similar recursion relations having a constant value of $\kappa$ and $\gamma$ are discussed by Lamura {\em et al.} \cite{Lamura:01}.

Examining Eq.~\ref{partition-function-high-F} we see that by integrating out the tangent vector degrees of freedom we have taken the Ising-model partition function corresponding to the secondary structure variables, which had only nearest-neighbor couplings, and transformed it into the partition function for the secondary structure variables ($s_i$) having interactions between the these variables at distant sites along the polymer chain. This result is to be expected: the combination of the coupling between the local chain tangents and secondary structure generated by $\kappa(s)$ combined with the long-range coupling of those chain tangents to each other over $\sim \kappa$ monomers leads to a new effective long-range interaction between secondary structure variables mediated by the conformational degrees of freedom of the chain.  It is clear from Eqs.~\ref{High-force-product}, \ref{R-def}, and \ref{R-def-0} that the simple Ising description of the secondary structure variables is recovered in limit of chains with a vanishing persistence length $\kappa \longrightarrow 0$ where the tangent vector degrees of freedom do not mediate a long-range interaction between the $s_i$ variables. In that case the recursion relation can be trivially solved to yield: $R_i = F \gamma(s_i)/2$ so that the remaining partition function of the secondary structure variables in Eq.~\ref{partition-function-high-F} reverts to that of an Ising model, but one for which each secondary structure configuration is weighted by its effect on the chain extension in the direction along the externally applied force. 

The short persistence length, decoupled limit is clearly not of primary interest in modelling an alpha-helical polypeptide. In fact, considering that we are primarily interested in molecules that are not significantly longer than their persistence length (in the alpha-helical phase), it appears physically reasonable to take a diametrically opposed approximation. For such chains where $\kappa_> \sim N$ one suspects that the statistics of the secondary structure variables is better represented in a mean field approximation enforced by the long-range interactions between these variables due to tangent vector correlations along the chain. To implement a mean field approach, we ignore boundary effects and study one secondary structure variable in the bulk of the chain, $s_i$.  This single degree of freedom interacts with the mean field of all the other secondary structure variables along the chain. We define the mean value of these variables as $m = \langle s_j \rangle,$ for all $j \neq i$. From this definition it is clear that $-1< m < 1$. In order to discuss the chain persistence length and effective monomer length we must generalize Eqs.~\ref{kappa-def},\ref{gamma-def} respectively by introducing the mean values of these quantities by defining
\begin{eqnarray}
\label{kappa-ave}
\bar{\kappa} = \frac{\kappa_{>}}{2} (1 + m) + \frac{\kappa_{<}}{2} (1 - m)  \\
\label{gamma-ave}
\bar{\gamma} = \frac{\gamma_{>}}{2} (1 - m) + \frac{\gamma_{<}}{2} (1 + m).
\end{eqnarray}
The linear dependence of these values on $m$ may be justified by noting that $\langle s_j \rangle = m$ implies that each segment spends a fraction $(m+1)/2$ of the time in its native state. At least on time scales long compared to interconversion time between the $s_j = \pm 1$ states, one would observe the effective values $\bar{\kappa}$ and $\bar{\gamma}$ as defined above.  Of course, nothing in the present analysis determines this interconversion time, but we expect it to be on the time of conformational changes of small molecules $\sim 10^{-9}$s. Both force spectroscopy measurements and protein conformational changes occur on much longer time scales where the approximation Eqs.~\ref{kappa-ave},\ref{gamma-ave} is valid.

We may write the mean field free energy of the chain under the externally applied force in the form
\begin{eqnarray}
\mathcal{F}_{\rm MF} = -   \log \left\{ \sum_{s_i = \pm 1} e^{-\frac{\epsilon_{\rm w}}{2}(m s_i - 1) + \frac{h}{2}(s_i-1)+ F \gamma(s_i)}  \right.  \nonumber \\
 \left. J_{\rm MF} \left[\bar{\kappa},\bar{\gamma},m;N/2-1 \right] \cdot  J[s_i,1] \cdot J_{\rm MF}  \left[\bar{\kappa},\bar{\gamma},m;N/2-1 \right] \right\}
. 
\label{MF-Free-Energy}
\end{eqnarray}
We have defined quantity $J_{\rm MF} \left[\kappa,\gamma,m;N \right]$ to be analogous to $J \left[\left\{s_i \right\}, N\right]$ (see Eq.~\ref{partition-function-high-F}) for a chain of $N$ monomers with a \emph{fixed} mean-field persistence length ($\bar{\kappa}$) and a \emph{fixed} mean-field monomer length ($\bar{\gamma}$).  We also take the mean field approximation: $s_k = m$ for all $k$ in the HC part of the Hamiltonian.  The function $J[s_i,1]$ represents the one Gaussian integral associated with angular degree of freedom at the $i^{\rm th}$ site. 

The physical meaning of Eq.~\ref{MF-Free-Energy} is that the free energy of the chain in the mean-field description may be written as the sum of three parts. The first part given by the negative logarithm of  $J_{\rm MF} \left[\bar{\kappa},\bar{\gamma},m;N/2-1 \right]$ gives the free energy of the the half of the chain to the left of the selected site $i$. This free energy is evaluated using the mean field approximation for the secondary structure variables and the Gaussian (small angle) approximation for the tangent vectors. The last term in the sum is the analogous contribution to the free energy associated with the length of polymer to the right of the selected site $i$ and evaluated using the same approximations. Finally, the middle term in the product appearing in Eq.~\ref{MF-Free-Energy} is the contribution to the free energy of the $i^{\rm th}$ site itself. The (Gaussian) integral $J[s_i,1]$ accounts for the tangent vector degree of freedom while the sum on one remaining $s_i$ variable is written explicitly above. 

To justify choosing the $i^{\rm th}$ site at the  middle of the chain to be representative of any site, we must ignore boundary effects. Consequently the mean-field description is most accurate in the limit of long chains, \emph{i.e.} $ \log N > \epsilon_{\rm w}$.  Finally, self-consistency requires that that the thermal average of the secondary structure at the $i^{\rm th}$ also be equal to $m$. Thus we demand 
\begin{equation}
\label{self-consistency}
m = \langle s_i \rangle =   2 \frac{\partial \mathcal{F}_{\rm MF}  }{\partial h}.
\end{equation}
From Eq.~\ref{self-consistency} we obtain a solution for $m$ that, when used in conjunction with the mean-field free energy function, $\mathcal{F}_{\rm MF} = - \log J_{\rm MF} \left[\bar{\kappa},\bar{\gamma},m;N \right]$ gives a complete thermodynamic description of the chain under an externally applied force. In particular we compute the mean length of the chain under these conditions from
\begin{equation}
\langle L \rangle = - \frac{\partial \mathcal{F}_{\rm MF} }{\partial F}.
\end{equation} 

\begin{figure}
\includegraphics[width=6cm]{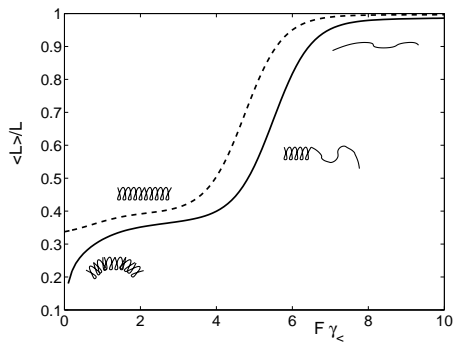}
\caption{\label{MF-length-fig} The mean length of the chain as a function of the applied force normalized by the maximum length $L$ computed using the mean field approximation discussed in the text. The parameters for the solid line at $\epsilon_{\rm w}=10$, and $h=1$, and  $\kappa_{>}= 2$ and $\kappa_{<}=1$ and $N=10$. The dashed line shows an analogous plot using parameter values of $h=1.5$, $\epsilon_{\rm w} = 8$, $\kappa_> =100$, and $\kappa_< =1$. These latter values are representative of alpha-helical protein domains.}
\end{figure}

We plot the mean extension of the polymer $\langle L \rangle$ in the direction of the applied force normalized by the maximal extension of the chain $L = N \gamma_>$ in figure~\ref{MF-length-fig}.  The applied force has been nondimensionalized by the length $\gamma_<$.  Qualitatively the figure may be discussed in terms of four regimes characterized by abrupt changes in the dependence of the mean length on applied force. For the smallest forces we see the initial extension of the predominant native-state, alpha-helical chain. As long as the $h > F \Delta \gamma$ the free energy decrease associated with the breakdown of the native state of molecule and the consequent extension of each monomer by $\Delta \gamma = \gamma_> - \gamma_<$ is more than off-set by the free energy increase per unit length of creating segments in this nonnative state, $h$. Thus the secondary structure variables are frozen in the native state and the extension of the chain proceeds by the suppression of contour fluctuations transverse to the extension direction. The effective maximal extension of the chain is $L \gamma_</\gamma_>$ (here $L/3$) and the saturation of chain extension reproduces the Marko, Siggia \cite{Marko:95} result so that $\langle L \rangle /L \sim  (1 - F^{-1/2})  N \gamma_</\gamma_> $.  

If the system were well-described by a WLC model, this high force plateau would be flat as the chain extension asymptotically approached it maximal value. The observed slow growth of the chain extension or ``pseudoplateau'' is due to the presence of local secondary structure fluctuations. With increasing force, these fluctuations are biased toward the more-extended, nonnative state so that the mean length of a monomer grows slowly with applied force. This second regime characterized by the pseudoplateau terminates in a rapid extension regime where the secondary structure is pulled apart by the applied force.  The requisite force to open up these alpha-helices determines the transition to this rapid extension regime. That force is given by $ F \Delta \gamma \sim \epsilon_{\rm w} + h$ where the net extension of one segment enthalpically compensates for the creation of a random coil segment and a domain wall on the chain.  Finally, in the fourth regime, the applied force has thoroughly destroyed the secondary structure. With increasing the force the now random-coil chain approaches its maximal extension $L$ in a manner first discussed by Marko and Siggia. 

\section{Conclusions}
\label{conclusions}

We have proposed and explored an extension of the traditional WLC polymer model in order to incorporate the presence of internal degrees of freedom along the polymer backbone and the coupling of those internal degrees of freedom to the conformational degrees of freedom of the chain. Such a model constitutes a minimal description of many biopolymer systems, but we focus on developing a description of an alpha-helical polypeptide chain. By studying the mechanical properties in thermal equilibrium of such polymers one will develop insight into the mechanical properties of {\em de novo} designed alpha-helical chains and  biomemtic synthetic polymers that acquire helical secondary structure in aqueous solution. 

Eventually one would like to apply such a model to entire proteins although such a program requires in general a mechanical description of beta-sheets and an investigation of the mechanical interaction of protein subdomains. In principle, the combination of a model of the nonlinear elastic properties of such domains with an accurate three-dimensional protein structure in its native state should enable the investigation of protein dynamics and particularly conformation change under biologically relevant conditions. It is known that at least some proteins are bistable having at least two structurally different conformations. In this paper we have shown that, due to the coupling of the conformational degrees of freedom to the internal, secondary structure variables, alpha helices are generically bistable mechanically. One may speculate as to whether this inherent bistability provides a mesoscopic mechanism to elucidate protein conformational change. 

In order to carry forward this program and to make quantitatively falsifiable predictions for the mechanical properties of single alpha helices it is necessary to determine the for energy scales that enter the HCWLC Hamiltonian.  Unfortunately, these four energy scales are imprecisely known at best. The better studied energy scales involve the helix coil (HC) parameters $\epsilon_{\rm w}$ and $h$. Both of these parameters are extremely difficult to estimate based on first principles since these energy scales involve complex solvation energies \cite{Chan:94} in addition to the formation of hydrogen bonds between adjacent turns of the alpha helix. These parameters can be estimated, however, by fitting HC models to both the results of molecular dynamics simulations and experiment \cite{Yang:95,Chakrabartty:94}. In terms of our parameters this work provides the following estimates: $\epsilon_{\rm w} \simeq 7$ and $h \simeq 1.5$. Thus we note that since typical alpha-helix domains in proteins have $N \sim \mathcal{O}(10)$ \cite{Rief:03,Emberly:03} these domains are highly cooperative, $\epsilon_{\rm w} > \log(N)$. 

There is little data on the persistence length of alpha helices. We estimate the bending modulus of an alpha helix by assuming that its enhanced stiffness arises primarily from inter-loop hydrogen bonding. Taking the energy scale of these hydrogen bonds to be $3 k_{\rm B} T$ -- $15 k_{\rm B} T$ \cite{Israelachvili:92} , an inter-loop distance of $0.36$nm and helical radius $\sim 0.1$nm we find that (with $k_{\rm B} T = 1$) the persistence length of the alpha-helix should be $\ell^{\rm h}_{\rm p} \sim 10$ - $50\mbox{nm}$. From this persistence length we determine the HCWLC bending modulus via: $\ell^{\rm h}_{\rm p}/\gamma_< = \kappa_{>} $ so that $\kappa_> \sim 25 - 140$. With the absence of hydrogen bonding in the nonnative, random coil state we assume that the persistence length is $ \ell_{\rm p}^{\rm c} \sim 1$nm, typical of simple hydrocarbons \cite{Doi:86}. Thus it is reasonable to suppose that $\ell_{\rm p}^{\rm c} /\gamma_> = \kappa_< \sim 3$. There is a significant dependence of thermal persistence length upon local secondary structure: the ratio $\kappa_>/\kappa_<$ may be as large as $50$. In order to explore the phenomenology of the model we have shown results for various parameter values, but we have always included plots corresponding to these biologically relevant parameters mentioned above. From these estimates of the biologically relevant energy scales in the model we predict the critical torques for the buckling failure of the alpha helix to be $\sim 40 \mbox{pN} \cdot \mbox{nm}$. The force required to pull out the alpha helices leading to the dramatic lengthening of the chain is roughly $150 - 200$pN. 

The central result of this paper is that an alpha-helical polypeptide is highly nonlinear in its response to applied forces and torques. The source of the nonlinearity is the coupling between local secondary structure and the conformational state of the polypeptide backbone. Bending or pulling on the alpha helix  mechanically can result in the abrupt breakdown of secondary structure and consequently a dramatic increase in bending and extensional compliance. The stresses required to access this highly nonlinear behavior occur on scales relevant to biological activity.  

\section*{Acknowledgements}

AJL would like to acknowledge F.\ Pincus, P.\ Nelson, K.A.\ Dill, T.C.B.\ McLeish, and R.\ Philips for helpful and enjoyable conversations. AJL also acknowledges the hospitality of the California Institute of Technology and the Newton Institute for Mathematical Sciences where some of this work was done. BC and AJL would also like to thank M. Muthukumar, R.A. Guyer, and J.F. Marko for stimulating conversations regarding this work.

\appendix
\section{Diagonalization of the Transfer Matrix}
\label{U-mat}
The transfer matrix at a given angular momentum $T(m)$ given in Eq.~\ref{T-mrep} is diagonalized by the similarity transformation: $D(m) = U^{-1}(m) \cdot T(m)\cdot  U(m)$ where the matrix $U(m)$ is
\begin{equation}
\label{U-matrix}
U(m) =
    \begin{pmatrix}
     \frac{\lambda_{1}(m) - 2 d_m}{2c_m}  &   \frac{\lambda_{2}(m) - 2 d_m}{2c_m}\\
     1        & 1
    \end{pmatrix},
\end{equation} 
where $\lambda_{1,2}(m)$ are the eigenvalues of the transfer matrix given in Eq.~\ref{eigenvalues} and the functions $c_m$ and $d_m$ are simply the bottom row of the transfer matrix. In other words
\begin{equation}
\label{c-def}
c_m = e^{-h - \epsilon_{\rm w} - \kappa_{<}} I_{m}[\kappa_{<}]
\end{equation}
and
\begin{equation}
\label{d-def}
d_m = e^{-h - \kappa_{<}} I_{m}[\kappa_{<}].
\end{equation}

\section{Correlation Function Coefficients}
\label{appendix}

Here we discuss the coefficients $A$ and $B$ appearing in Eq.~\ref{correlation-function-II}. These coefficients, which have dimensions of length squared, can be computed in terms of the matrices $\Gamma^{(01)}$, $\Gamma^{(10)}$ defined in Eqs.~\ref{Gamma-01} ,\ref{Gamma-10}; by calculating the necessary trace, one finds that
\begin{eqnarray}
A &=& \Gamma^{(01)}_{11} \Gamma^{(10)}_{11} \\
B &=& \Gamma^{(01)}_{12} \Gamma^{(10)}_{21}.
\end{eqnarray}
Determining these coefficients in terms of the fundamental parameters of the model is now a matter of some algebra. To simplify this work and to better display the result, we find it helpful to write $A$ and $B$ in terms of $\gamma_{<,>}$ and the transfer matrix eigenvalues $\lambda_{1,2}(m)$ given by Eq.~\ref{eigenvalues}. We find
\begin{eqnarray}
\label{A-definition}
A = \frac{4 c_0 c_1 \gamma_{>} - \gamma_{<}(2 d_1 - \lambda_1(1))(2 d_0 - \lambda_2(0)) }{4 c^{2}_0 - (2 d_0 - \lambda_1(0))(2 d_0 - \lambda_2(0))} \cdot \\
\frac{4 c_0 c_1 \gamma_{>} - \gamma_{<}(2 d_0 - \lambda_1(0))(2 d_1 - \lambda_2(1))}{4 c^{2}_1 - (2 d_1 - \lambda_1(1))(2 d_1 - \lambda_2(1))} \nonumber
\end{eqnarray}
and
\begin{eqnarray}
\label{B-definition}
B = \frac{4 c_1 \gamma_{<}(2 d_0 - \lambda_2(0)) + c_0 \gamma_{>}(-2 d_1 + \lambda_2(1))}{4 c^{2}_0 - (2 d_0 - \lambda_1(0))(2 d_0 - \lambda_2(0))} \cdot \\
\frac{c_1 \gamma_{<}(-2 d_0+ \lambda_1(0)) + c_0 \gamma_{>}(2 d_1 - \lambda_1(1))}{4 c^{2}_1 - (2 d_1 - \lambda_1(1))(2 d_1 - \lambda_2(1))} \nonumber
\end{eqnarray}
In the above equations we have used the functions defined in Eqs.~\ref{c-def},\ref{d-def}.

\end{document}